\documentstyle[psfig]{elsart}

\def\lcdm{LCDM}
\newcommand{\hmpc}{\mbox{$h^{-1}$ Mpc}}
\def\la{\mathrel{\hbox{\rlap{\hbox{\lower4pt\hbox{$\sim$}}}\hbox{$<$}}}}
\def\ga{\mathrel{\hbox{\rlap{\hbox{\lower4pt\hbox{$\sim$}}}\hbox{$>$}}}}
\def\gsim{\ga}
\def\lsim{\la}

\begin{document}

\begin{frontmatter}
\title{Probing Galaxy Formation with TeV Gamma Ray Absorption}

\author[1,2]{Joel R. Primack}
\author[1]{James S. Bullock}
\author[2]{Rachel S. Somerville}
\author[3]{Donn MacMinn}

\address[1]{Physics Department, University of California, Santa Cruz,
CA 95046 USA}
\address[2]{Racah Institute of Physics, The Hebrew University,
Jerusalem 91904, Israel}
\address[3]{Deceased}

\begin{abstract}
We present here the extragalactic background light (EBL) predicted by
semi-analytic models of galaxy formation, and show how measurements of the
absorption of gamma rays of $\sim$ TeV energies via pair production on the EBL
can probe cosmology and the formation of galaxies.  Semi-analytic
models permit a physical treatment of the key processes of galaxy formation --
including gravitational collapse and merging of dark matter halos, gas cooling
and dissipation, star formation, supernova feedback and metal production -- and
have been shown to reproduce key observations at low and high redshift. Using
this approach, we investigate the consequences of variations in input
assumptions such as the stellar initial mass function and the underlying
cosmology. We conclude that observational studies of the absorption of
$\sim 10^{-2}-10^{2}$ TeV
gamma rays will help to constrain the star formation history of the universe,
and the nature and extent of the extinction of starlight due to dust and
reradiation of the absorbed energy at infrared wavelengths.
\end{abstract}

\end{frontmatter}


\section{Introduction}

It has long been appreciated that high energy $\gamma$-rays from sources at
cosmological distances will be absorbed via electron-positron pair production
\cite{gould67} on the diffuse background of long wavelength photons produced
over the history of the universe. Now that several extragalactic TeV sources
have been discovered, it is beginning to become possible to use this process to
probe the extragalactic background light (EBL). This technique will become
increasingly powerful as many more sources will presumably be discovered at
greater distances with the new generation of $\gamma$-ray telescopes (GLAST,
CELESTE, STACEE, MAGIC, HESS, VERITAS, and Milagro).

Broadly speaking, there are three approaches to studying the
EBL/$\gamma$-absorption connection, represented by the three speakers
on this topic here at the VERITAS workshop:
\begin{itemize}
\item Limits on the EBL, and on models for its production, from
$\gamma$-absorption data \cite{biller};
\item Semi-empirical estimates of the EBL \cite{stecker}; and
\item Prediction of the EBL and $\gamma$-absorption from physical
theories of galaxy formation and evolution in a cosmological framework.
\end{itemize}
The advantage of the last of these approaches, which we will follow in this
talk, is that it permits one to deduce from $\gamma$-ray absorption data a
great deal about galaxy formation and evolution, including the effects of the
stellar initial mass distribution and of dust extinction and reradiation.  It
is also arguably the best way to estimate the extent of $\gamma$-ray absorption
at various energies, as shown by the correct prediction from a simplified model
of this sort \cite[hereafter MP96]{mp} that there would be rather little
absorption of TeV $\gamma$-rays from the nearest extragalactic sources, Mrk 421
and Mrk 501, at redshifts of only $z=0.03$.  The calculations reported here
(and in more detail in \cite{sbp} and \cite{bsmp}) are based on
state-of-the-art semi-analytic models (SAMs) of galaxy formation, which we
summarize briefly below.  But it will be useful to start by summarizing our
earlier calculations (MP96).

\section{Simplified Cosmological Modeling of the EBL}
Although there is a long history of spectral synthesis models leading to
predictions for the EBL (e.g., \cite{pp67,yt88,franceschini94}), such models
typically attempted to account only for the star formation history of the
galaxies existing today --- i.e., they are pure luminosity evolution models.
Moreover in spectral synthesis, 
the star formation history of each galaxy is {\it not} determined from its
cosmological history, taking into account the fact that gas is not available to
form stars until it has cooled within a dense collapsed structure. But the
evidence is certainly increasing that there was a great deal of galaxy
formation and merging in the past, plausibly in agreement with the predictions
of hierarchical models of galaxy formation of the CDM type.

The motivation of the approach used by MP96 was to obtain theoretical
predictions for the EBL and the resulting absorption of $\gamma$-rays in the
context of hierarchical theories of structure formation, specifically within the
CDM family of cosmological models. It is relatively straightforward to
calculate the evolution of structure in the dark matter component within the
CDM paradigm. In MP96, the number density of dark matter halos as a
function of mass and redshift and its dependence on cosmology was modeled using
Press-Schechter theory, which agrees fairly well with the predictions of N-body
simulations. However, obtaining the corresponding radiation field as a function
of time and wavelength involves complicated astrophysics with many unknown
parameters. This problem was addressed empirically, using the simple assumption
that each dark matter halo hosts one galaxy, with the galaxy luminosity assumed
to be a monotonic function of the halo mass. To model the spectrum of each
galaxy, the star formation rate (SFR) was assumed to be $\propto e^{-t/\tau}$,
with ellipticals (assumed to be 28\% of the galaxies) having $\tau=0.5$ Gyr,
and spirals (the remaining 72\%) having $\tau=6$ Gyr. The actual SFR for each
galaxy was determined so that a desired local luminosity function (LLF) was
reproduced at redshift $z=0$. As there is considerable variation in the
observed B-band luminosity function derived from different redshift surveys,
MP96 considered three representative choices.

Stellar emission was modeled in a simplified way, with each stellar population
of a given mass and age treated as a black body with the appropriate
temperature. Three different power-law initial mass functions (IMFs) $N(M)
\propto (M/M_\odot)^{-1+\Gamma}$ describing the differential distribution of
the initial stellar masses were considered: the standard Salpeter IMF with
$\Gamma=-1.35$, and also steeper IMFs with $\Gamma=-1.6$ and -2.0. The
evolution of the gas content and metallicity within each galaxy was treated in
the instantaneous recycling approximation. Dust absorption was treated in a
standard way \cite{GuiderdoniRV}, assuming that the mass of dust increases with
the galaxy metallicity and gas fraction. The extinction curve was similar to a
Galactic extinction curve, but was scaled according to the metallicity
(galaxies with lower metallicities have steeper extinction curves in the UV, as
indicated by observations of the LMC and SMC).  Energy was conserved, so that
any energy absorbed by dust was reradiated. The dust emission spectrum was
modeled \cite{desert} with three components: PAH molecules ($\sim 10$ to 30
$\mu$m), warm dust from active star forming regions (30 to 70 $\mu$m), and cold
``cirrus'' dust (70 to 1000 $\mu$m).

Two cosmological models were considered, a standard (cluster-normalized,
$\sigma_8=0.67$) $\Omega_{\rm matter}=1$ cold dark matter (SCDM) model, and a
COBE-normalized cold + hot dark matter (CHDM) model with the then-favored hot
dark matter fraction $\Omega_\nu=0.3$. Since both were $\Omega_{\rm matter}=1$
models, the Hubble parameter was chosen to be $h=0.5$ ($H_0=100 h$ km s$^{-1}$
Mpc$^{-1}$) in order to obtain a Universe with an age of 13 Gyr. Galaxy
formation occurs fairly early in the SCDM model (because of the large amount of
power on small scales), and considerably more recently in the CHDM model.

The main conclusion from this study was that cosmology is the dominant factor
influencing the EBL in the range 1-10 $\mu$m, which is the range most relevant
for absorption of $\sim$ TeV $\gamma$-rays from nearby extragalactic
sources. In this wavelength range, the most extreme differences between the
three different LLFs and three different IMFs considered were less than the
difference between the SCDM and CHDM cosmological models, representing early
and late galaxy formation respectively. It is not difficult to understand why
this happens: since the optical luminosity at $z=0$ was fixed for each assumed
LLF, the main factor determining the EBL in the near infrared was the star
formation history. Because galaxies were assumed to trace halos in a simple
way, the star formation history was almost entirely determined by the
cosmology. As MP96 explained, the SCDM model predicted a larger EBL flux
because (1) the stars have put out more light since they have been shining
longer than in CHDM, (2) there was more redshifting of their light from the
optical to the near infrared, and (3) the SCDM galaxies are older at a given
redshift and hence are composed of more evolved stars, producing more flux in
the red and near-infrared.

As expected, when the optical depth for $\gamma$-rays due to $e^+e^-$
production was calculated, more absorption was predicted for SCDM than for
CHDM.  But for sources as near as Mrk 421 and 501, the predicted absorption
only steepens the spectrum a little in the 300 GeV - 10 TeV range for which
results have thus far been published, with curvature noticeable mainly above
about 3 TeV. These predictions appear to be consistent with the observations
\cite{zweerink}, unlike those from earlier \cite{stecker94} and later
\cite{stecker97} calculations based on a semi-empirical approach. The
predictions of our new, more complete treatment are qualitatively consistent
with this earlier simplified approach.

\section{Semi-Analytic Modeling of the EBL}
Our new approach is based on semi-analytic models (SAMs) of galaxy formation,
which allow one to model the astrophysical processes involved in galaxy
formation in a simplified but physical way within the framework of the
hierarchical structure formation paradigm. The semi-analytic models used here
are described in detail in \cite{rachelthesis}, \cite{sp98}, and
\cite{spf98}. These models are in reasonably good agreement with a broad range
of local galaxy observations, including the relation between luminosity and
circular velocity (the Tully-Fisher relation),
the B-band luminosity function, cold gas
contents, metallicities, and colors \cite{sp98}. Our basic approach is similar
in spirit to the models originally presented by \cite{kwg93} and
\cite{cole94}, and subsequently developed by these groups in numerous other
papers (reviewed in \cite{rachelthesis} and \cite{sp98}). Significant
improvements included in \cite{sp98} are that we assumed a lower stellar
mass-to-light ratio (in better agreement with observed values), included the
effects of dust extinction, and developed an improved ``disk-halo'' model for
supernovae feedback. With these new ingredients, we were able to overcome some
of the difficulties of previous models, which did not simultaneously reproduce
the Tully-Fisher relation and B-band luminosity function, and produced bright
galaxies that were too blue.

\begin{figure}
\centerline{\psfig{file=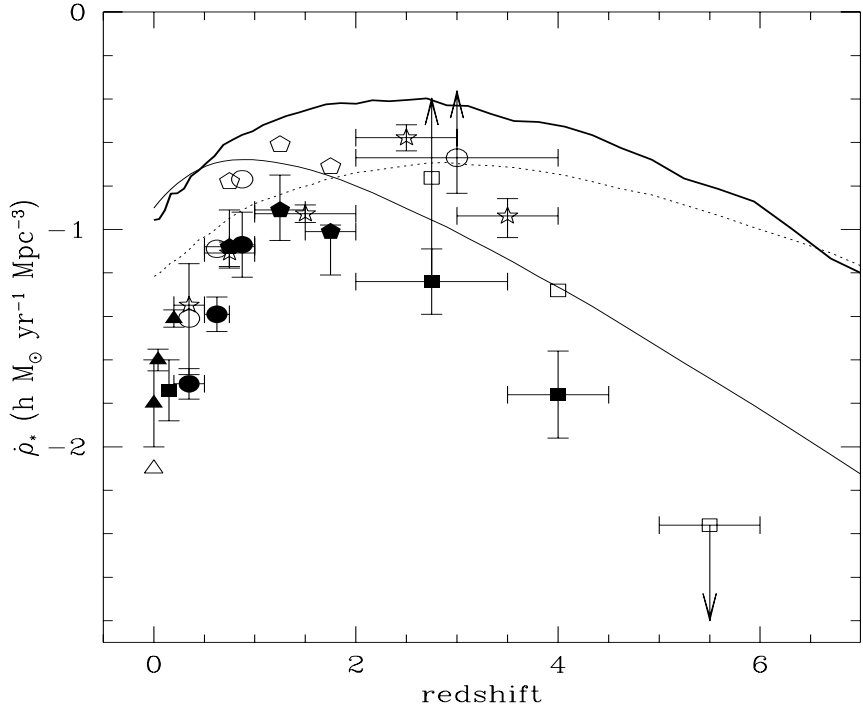,width=12truecm,height=10truecm}}
\caption{The global star formation rate as a function of redshift, for three
different SCDM models of star formation. The light solid line shows a model
with constant star formation efficiency, including quiescent star formation
only. The bold solid line assumes constant efficiency quiescent star formation,
but contains a contribution from starbursts induced by galaxy-galaxy collisions
\protect\cite{spf98}.  The dotted line represents only quiescent star
formation, but the efficiency of star formation is assumed to scale inversely
with the dynamical time of the galaxy, therefore star formation is more
efficient at high redshift because the typical objects are smaller and have
shorter dynamical times. This model is similar to the fiducial model of
Somerville \& Primack \protect\cite{sp98}, and to the models of Kauffmann et
al.  The data shown as filled symbols are uncorrected for extinction; the
corresponding open symbols show the extinction correction suggested in
\protect\cite{pettini:dust}.  The tip of the left vertical arrow shows a
correction of a factor of seven, which may be indicated by recent observational
studies of dust extinction in Lyman break galaxies.  The large open circle at
$z=3$ shows the lower limit from the SCUBA observations of the HDF region
\protect\cite{scuba} (assuming that these sources are stellar rather than AGN).
(More complete descriptions of the data with references are given in
\protect\cite{spblois}.)  Note that the discrepancy between the predictions and
the data at low redshifts ($z\lsim1$) is partly alleviated when the models are
corrected downward by about a factor of 2, to correct for the overprediction of
the number density of $M\lsim 0.1 M_*$ halos in the Press-Schechter
approximation.}
\label{fig:sfrz}
\end{figure}

Instead of assuming a one-to-one relationship between galaxies and dark matter
halos, as in MP96, we now determine the galaxy population residing in halos of a
given mass by constructing the ``merging history'' of each halo using an
extension of the Press-Schechter technique. Using the method described in
\cite{sk98}, we create Monte-Carlo realizations of the masses of progenitor
halos and the redshifts at which they merge to form a larger halo. These
``merger trees'' (each branch in the tree represents a halo merging event)
reflect the collapse and merging of dark matter halos within a specific
cosmology and have been shown to agree fairly well with merger trees extracted
from N-body simulations \cite{slkd}. Each halo at the top level of the
hierarchy is assumed to be filled with hot gas, which cools radiatively and
collapses to form a gaseous disk. The cooling rate is calculated from the
density, metallicity, and temperature of the gas. Cold gas is turned into stars
using a simple recipe, depending on the mass of cold gas present and the
dynamical time of the disk. Supernovae inject energy into the cold gas and may
expell it from the disk and/or halo if this energy is larger than the escape
velocity of the system. Chemical evolution is traced assuming a constant yield
of metals per unit mass of new stars formed. The spectral energy distribution
(SED) of each galaxy is then obtained by assuming an IMF and using stellar
population models (e.g. \cite{bc93}; in the present work we use the updated
GISSEL98 models with solar metallicity). When halos merge, the galaxies
contained in each progenitor halo retain their seperate identities until they
either fall to the center of the halo due to dynamical friction and merge with
the central galaxy, or until they experience a binding merger with another
satellite galaxy orbiting within the same halo. All galaxies are assumed to
start out as disks, and major (nearly equal mass) mergers result in the
formation of a spheriod. New gas accretion and star formation may later form a
new disk, resulting in a variety of bulge-to-disk ratios at late times. This
may be used to divide galaxies into rough morphological types, and seems to
reproduce observational trends such as the morphology-density relation and
color-morphology trend \cite{kwg93,bcf:96}.

The recipes for star formation, feedback, and chemical evolution contain free
parameters, which we set by requiring an average fiducial ``Milky Way'' galaxy
to have an I-band magnitude, cold gas mass, and metallicity as dictated by
observations of nearby galaxies. The star formation and feedback processes are
some of the most uncertain elements of these models, and indeed of any attempt
to model galaxy formation. We have investigated several different combinations
of recipes for star formation and supernova feedback (sf/fb), discussed in
detail in \cite{sp98}, \cite{spf98} and \cite{spblois}. The star formation
history for several different scenarios in shown in Figure~\ref{fig:sfrz}. Note
that the three models shown in Figure~\ref{fig:sfrz} are for the same
SCDM cosmology;
the only difference is in the mechanism used to convert cold gas
into stars. This illustrates that, unlike in the previous approach of
MP96, the star formation history of the Universe is quite sensitive to the
assumed astrophysics and not only the cosmology. Here we will discuss results
for a single choice of sf/fb recipe, which corresponds to the fiducial ``Santa
Cruz'' model discussed in \cite{sp98}, and is similar to the models of
Kauffmann et al. (e.g. \cite{kwg93,kauffmann98}). We shall elaborate on the
effects of changing the sf/fb recipes on the EBL and gamma ray absorption in
\cite{sbp} and \cite{bsmp}.

In \cite{sp98}, we included dust extinction using the same approach as
MP96.  As shown in \cite{sp98}, this led to better results for the B-band
luminosity function, and improved galaxy colors. Using a similar approach to
modelling dust extinction, \cite{kauffmann98} confirmed these results and
demonstrated that in addition the inclusion of dust greatly improves the
agreement of the galaxy-galaxy correlation function with observations. The
inclusion of dust extinction and the re-radiation of absorbed light at longer
wavelengths is of course a crucial ingredient in modeling the EBL. The current
dust model, discussed in \cite{rachelthesis}, is an improved version of the one
used in MP96, and is very similar to the approach used by Guiderdoni et
al. \cite{Guiderdoni98}. As in MP96, all of the absorbed starlight is
re-radiated by the dust, assuming a three-component blackbody emission
spectrum. However, in MP96 the shape of the dust emission spectrum was
chosen to match that of the Galaxy, which is inconsistent with the global data
for IRAS galaxies. In the current models, the temperatures and relative
contributions of the three components are determined by requiring the colors at
12, 25, and 60 $\mu$m to match the observed colors of IRAS galaxies
\cite{soifer:87}. Details will be given in \cite{sbp}.

\section{Initial Mass Function}

\begin{figure}
\centerline{\psfig{file=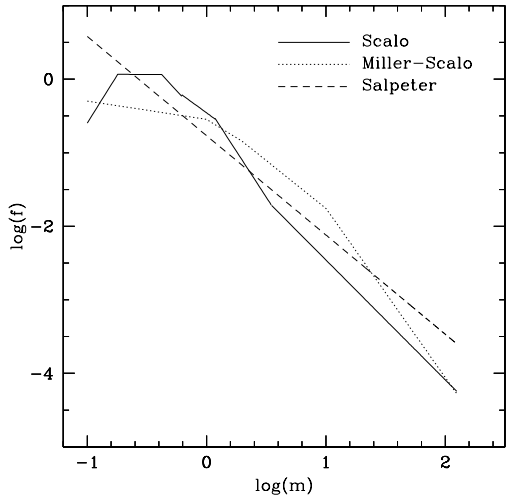,height=7truecm}}
\caption{Stellar initial mass functions (IMFs): Scalo
\protect\cite{scalo86}, Miller-Scalo \protect\cite{millerscalo}, and
Salpeter \protect\cite{salpeter}; $f(m)dm$ is the mass fraction in
stars with mass in the interval $m,m+dm$, where $m$ is in units of
$M_\odot$.}
\label{fig:IMFs}
\end{figure}

\begin{figure}
{\noindent \begin{minipage}[t]{2.5in} \centering 
 {\psfig{file=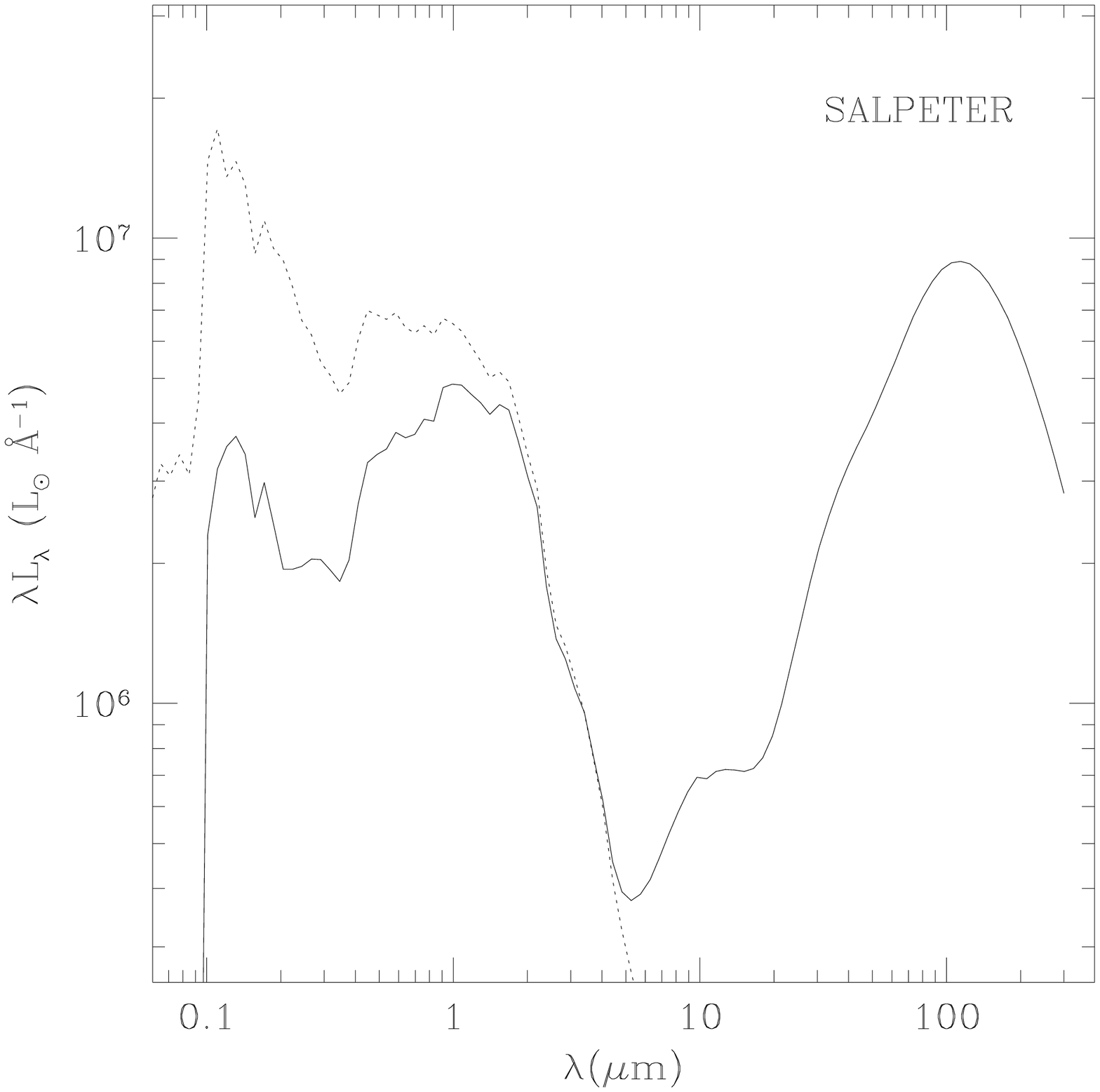,height=2.5in}} \end{minipage} \hfill
 \begin{minipage}[t]{2.5in} \centering 
 {\psfig{file=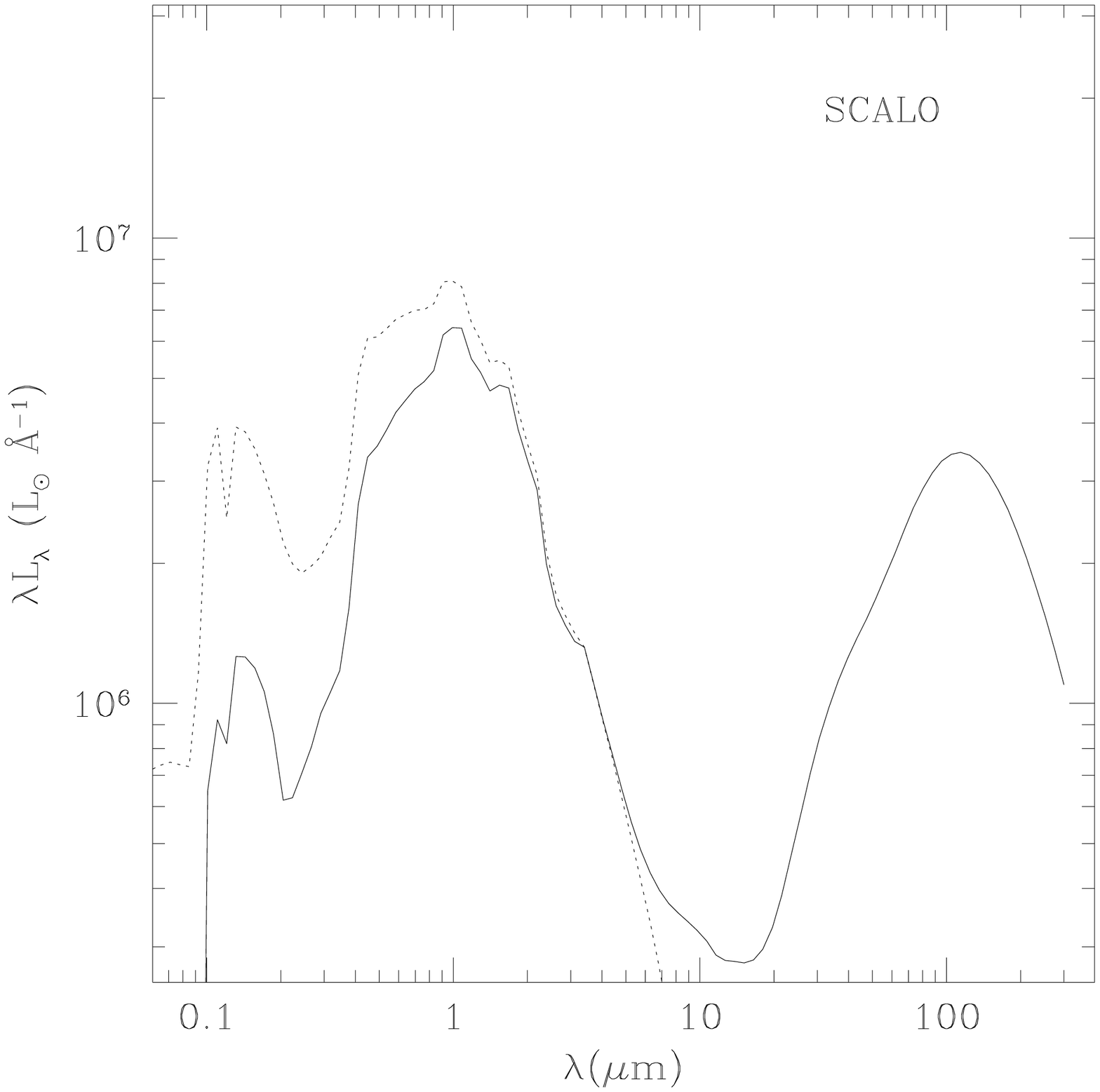,height=2.5in}} \end{minipage} }
\caption{The spectrum of an average ``Milky Way'' sized galaxy,
assuming a Salpeter (left) or Scalo (right) IMF. The dotted lines
show the spectra of starlight without the effects of dust. The solid
lines show the spectra with the effects of extinction and emission by
dust.}
\label{fig:galspectra}
\end{figure}

The stellar initial mass function determines the wavelength distribution of
starlight produced by a stellar population of a given age, and as such it is an
important ingredient in the calculation of the EBL. In MP96 we considered only
the Salpeter IMF and steeper power-law IMFs, but here we will discuss results
both for the Salpeter IMF and the Scalo IMF (see Figure~\ref{fig:IMFs}).
These are two of
the most commonly used IMFs, although recent studies (e.g., \cite{scalo98})
indicate that a better representation of the observed IMF may be a
Salpeter-like slope at $M \gsim M_\odot$, with a flattening at $M \lsim
M_\odot$.  The most important difference between the Salpeter and Scalo IMFs
for our present purposes is that if both are normalized to the same total mass
of stars, the fraction of high-mass stars is higher with the Salpeter IMF.
Since only high-mass stars emit significant amounts of ultraviolet light, this
results in much more ultraviolet in the spectrum of a typical galaxy with
Salpeter IMF as compared with Scalo IMF, as shown in Figure~\ref{fig:galspectra},
both without and
with inclusion of the effects of dust.  Note that the much greater amount of
absorbed ultraviolet light with the Salpeter IMF results in much more
reradiated infrared light at long wavelengths.  Also note that changing the IMF
has a relatively small effect on the predictions in the 1-10 $\mu$m range.

\begin{table*}
\caption{Parameters of Cosmological Models. From left to right, the
tabulated quantities are: the age of the universe in Gyr, the Hubble
parameter, the matter density, baryon density, density of hot dark
matter, density in the form of a cosmological constant in units of
the critical density, the slope (or `tilt') 
of the primordial power spectrum, and
the rms mass variance on a scale of $8\hmpc$. The last two columns
indicate whether the models are consistent with the observed cluster
abundance and the COBE normalization, respectively.}
\begin{center}
\begin{tabular}{lllllllllcc}
\hline
Model& $t_0$& $h$ & $\Omega_m$ & $\Omega_b$& $\Omega_{\nu}$&
$\Omega_\Lambda$ &
n & $\sigma_8$ & clusters & COBE\\
\hline
SCDM         &13.0 & 0.50 &1.0 &0.08 &0.0 &0.0 & 1.0 & 0.67 & Y & N \\
CHDM         &13.0 & 0.50 &1.0 &0.08 &0.2 &0.0 & 1.0 & 0.65 & Y & Y \\
OCDM         &12.7 & 0.60 &0.5 &0.056 &0.0 &0.0 & 1.0 & 0.85 & Y & Y \\
LCDM         &14.5 & 0.60 &0.4 &0.056 &0.0 &0.6& 1.0 & 0.84 & Y & Y \\
\hline
\end{tabular}
\end{center}
\label{tab:cosmo}
\end{table*}

\begin{figure}
\centerline{\psfig{file=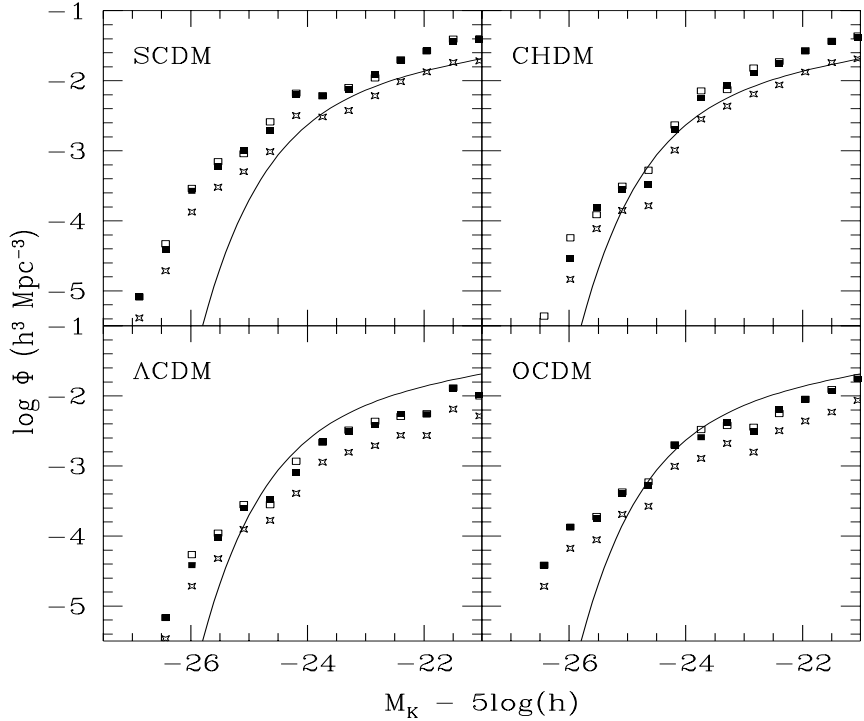,width=12truecm,height=10truecm}}
\caption{The predicted K-band luminosity function for the four cosmologies
considered here. Open and filled squares indicate the model results with and
without dust extinction, respectively, and the star symbols indicate the
approximate correction for the inaccuracy of the Press-Schechter method (see
text). Solid lines show fits to the recent observational determination of the
K-band luminosity function \protect\cite{szokoly}.
}
\label{fig:lfk}
\end{figure}

\begin{figure}
\centerline{\psfig{file=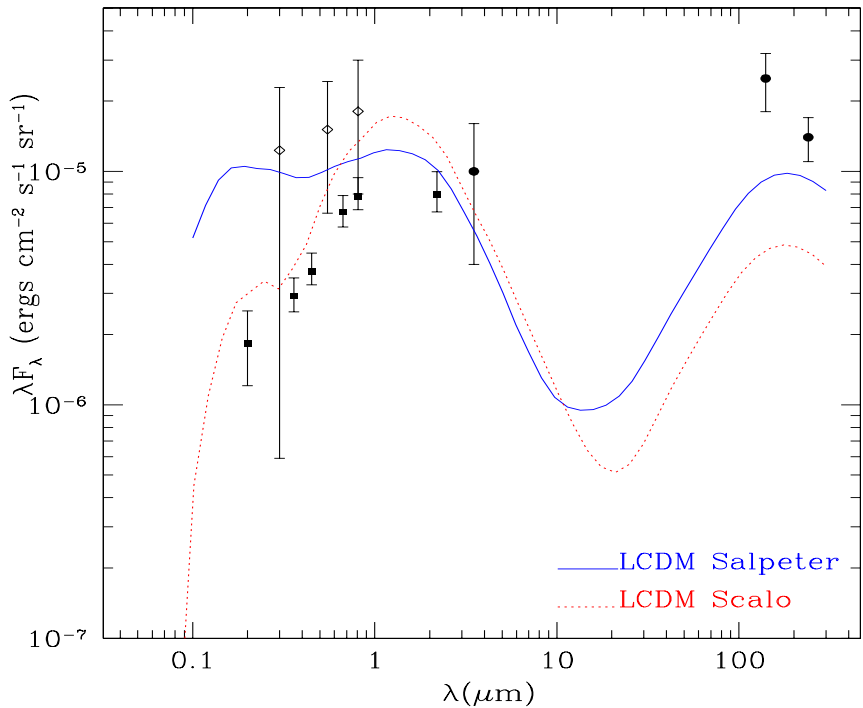,width=12truecm,height=10truecm}}
\caption{The $z=0$ EBL as calculated using SAM models for an LCDM
universe assuming both Scalo and Salpeter IMFs.  The Salpeter IMF
produces more high mass stars, thus more ultraviolet light to be
absorbed and re-radiated by dust at $\sim 100$ $\mu$m.  The filled
dots are DIRBE detections of the EBL at 140 and 240 $\mu$m
\protect\cite{dirbeEBL} and 3.5 $\mu$m \protect\cite{DwekArendt98}.
The solid squares are from \protect\cite{Pozzetti98} using HDF galaxy
counts, and the open diamonds are from \protect\cite{Bernstein98}, as
reported by \protect\cite{Dwek98}.}
\label{fig:lcdmSalScalo}
\end{figure}

\begin{figure}
\centerline{\psfig{file=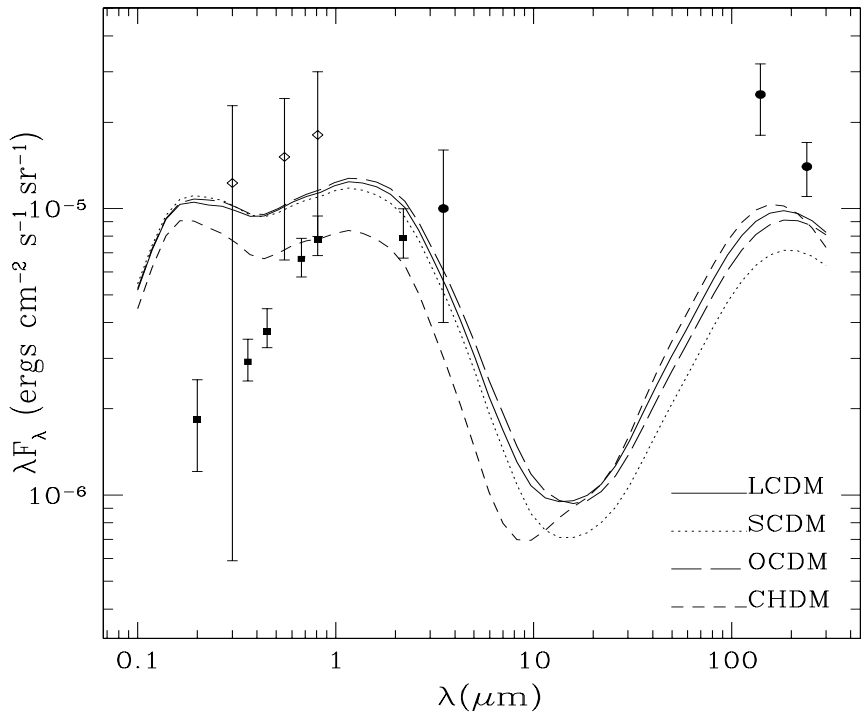,width=12truecm,height=10truecm}}
\caption{Predicted extragalactic background light for all four
cosmologies considered here (Salpeter IMF).
The data points are the same as those
in Figure~\ref{fig:lcdmSalScalo}.  The three models in which galaxy and
star formation is relatively
early predict similar EBL, while CHDM predicts
lower EBL, because galaxies form more recently.}
\label{fig:ebl_all}
\end{figure}

\begin{figure}
\centerline{\psfig{file=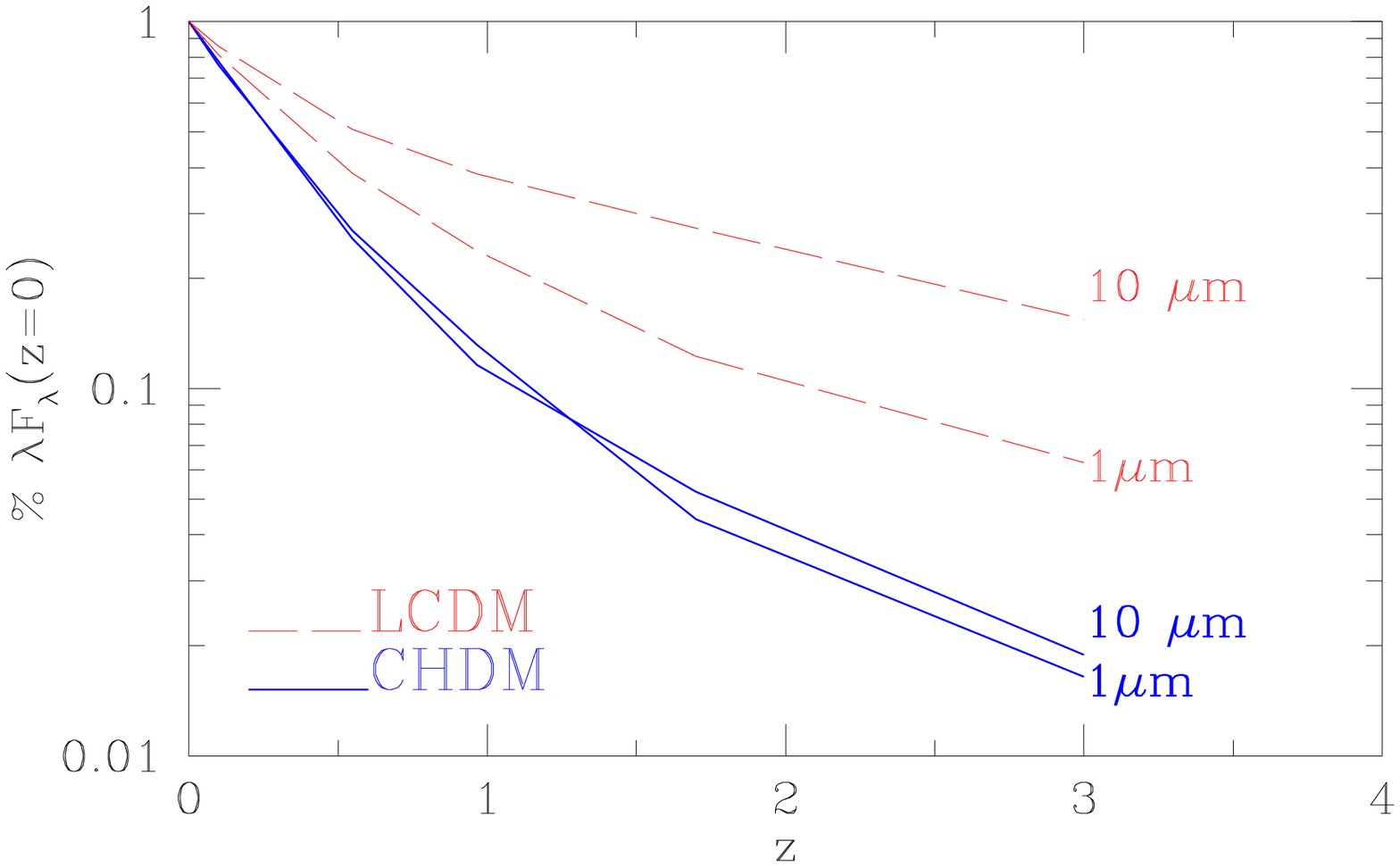,width=12truecm,height=10truecm}}
\caption{Evolution of the extragalactic background light.  Each curve
shows the fraction of the 1 and 10 $\mu$m
EBL ($\lambda F_{\lambda}$) at the present epoch ($z=0$)
coming from galaxies radiating
at redshift $\geq z$.  We show the evolution for our CHDM
and LCDM models.  Due to
the later onset of structure formation in the CHDM model relative
to LCDM, a much lower fraction of the $z=0$ EBL comes from galaxies at
high-redshift.}
\label{fig:evol}
\end{figure}

Previous SAM calculations of properties of local galaxies
(e.g. \cite{kwg93,cole94,sp98}) assumed a Scalo IMF. However, in \cite{baugh98}
and \cite{spf98}, it was shown that a more ``top-heavy'' IMF (more high-mass
stars) such as the Salpeter IMF is favored by observations of UV-bright
galaxies at very high redshift (Lyman-break galaxies). In \cite{sbp}, we
investigate the effects of using different IMFs (Scalo or Salpeter) on the
observable properties of galaxies at $z=0$. In particular, we calculate the
luminosity function at 2000\AA, B, R, and K, the corresponding 2000\AA-B, B-V,
B-I, and B-K colors predicted by our models, and the Tully-Fisher relation in
various bands. We find that because the mass-to-light ratio in the longer
wavebands (I to K) is significantly higher for the Salpeter IMF, the luminosity
of a galaxy with a given velocity dispersion is smaller. This makes it very
difficult to obtain a bright enough Tully-Fisher zero-point in cosmologies with
$\Omega_{\rm matter}=1$, in which the baryon fraction is low (we assume a fixed
value of $\Omega_b h^{2} = 0.02$, as suggested by observations
\cite{baryons}; the baryon fraction $f_b \equiv \Omega_b/\Omega_{\rm
matter}$ is therefore higher in low-$\Omega_{\rm matter}$ cosmologies).
If the mass-to-light ratios predicted by
the current generation of stellar population models are accurate and the
Salpeter IMF is really representative of typical galaxies, this may suggest
that the baryon fraction in bright galaxies must be $0.15-0.2$, similar to the
value in groups and clusters.

\section{Predicted EBL}

In \cite{sbp}, we calculate the predicted EBL for several cosmologies and also
investigate the effects of the assumed IMF and star formation recipe, and the
variations among stellar population models compiled by different groups. Here
we can only present a subset of preliminary results. The cosmological models
considered are summarized in Table 1. The SCDM model is shown for comparison
with other work, but is ruled out by many independent observational
considerations. The three remaining models represent currently favored variants
of the CDM family of models. The shape and normalization of the luminosity
function that we obtain from the SAMs depends on the cosmological model, as
shown in Figure~\ref{fig:lfk}. The K-band luminosity function shown in the
figure is relatively insensitive to dust and the star formation history. To put
the models on an equal footing, following a similar logic to the
B-band local luminosity function normalization of
MP96, we renormalize each model to give the same integrated luminosity in
the K-band. The common normalization is obtained by integrating the observed
local
luminosity function from Ref.~\cite{szokoly}. The resulting correction factors
range from 0.42 for SCDM to 1.7 for LCDM. Note that this renormalization also
sidesteps the known inaccuracy of the Press-Schechter model used to estimate
the number density of dark matter halos. The factor of 0.42 for SCDM is
consistent with the rule-of-thumb factor of 0.5 determined from comparison with
N-body simulations \cite{slkd,gross98}.

\begin{figure}
\centerline{\psfig{file=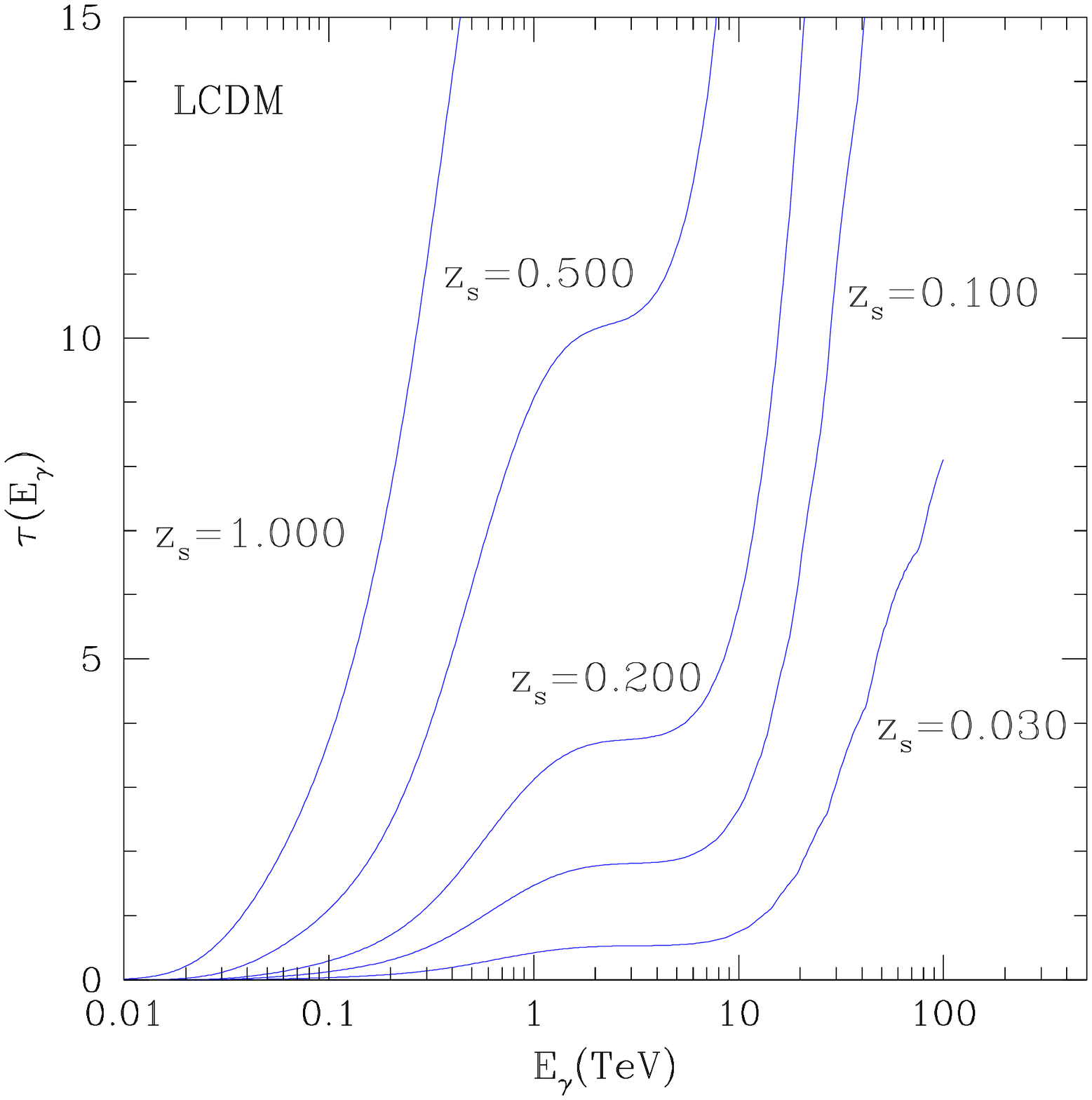,width=12truecm,height=10truecm}}
\caption{Optical depth of the universe to $\gamma$-rays as a function
of $\gamma$-ray energy, $E_{\gamma}$, for \lcdm\ with Salpeter IMF.
The assumed redshift of the source, z$_s$, is indicated for each curve. }
\label{fig:tau}
\end{figure}

\begin{figure}
\centerline{\psfig{file=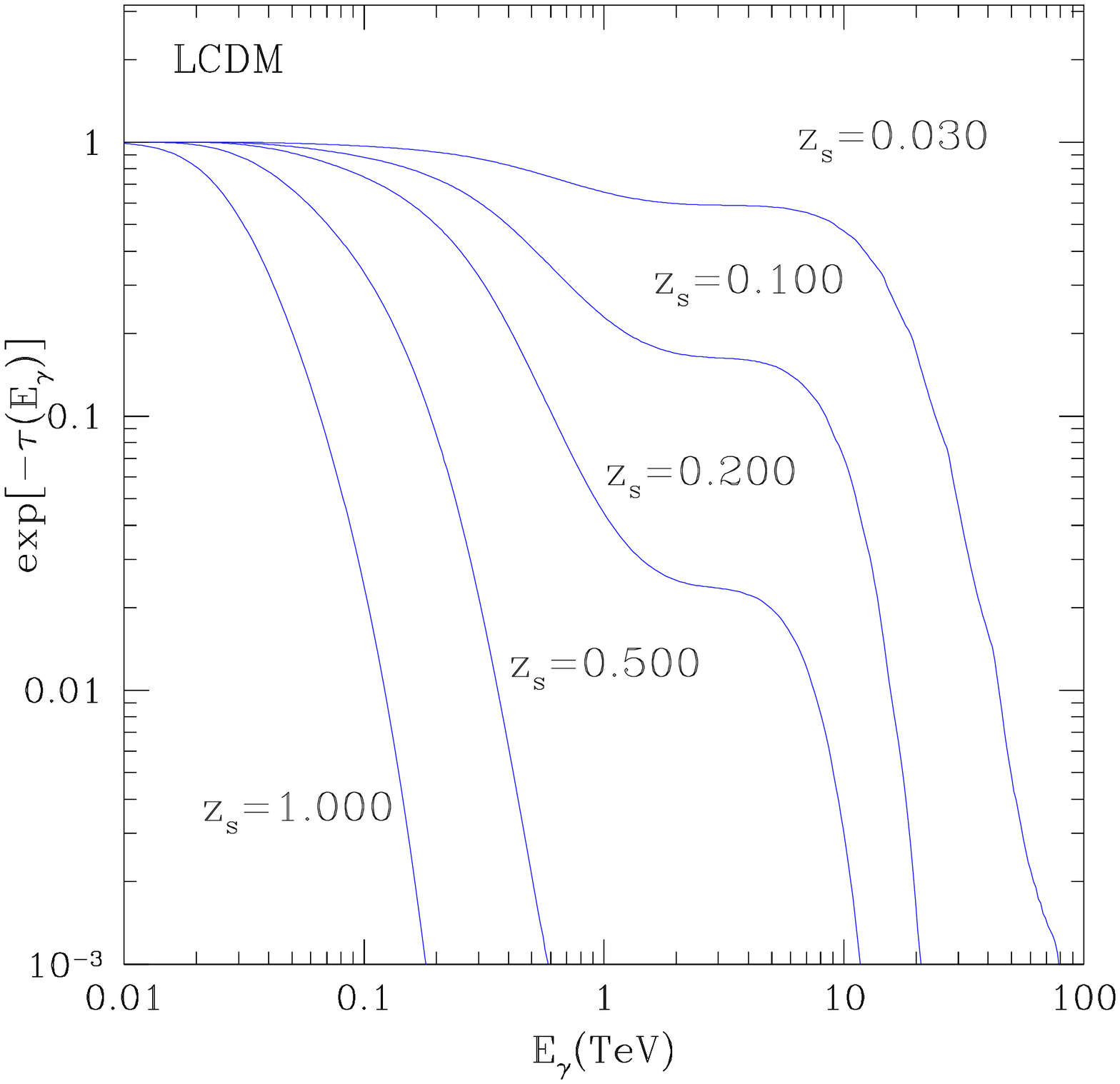,width=12truecm,height=10truecm}}
\caption{The attenuation factor, $\exp(-\tau)$ for $\gamma$-rays
as a function of $\gamma$-ray
energy for \lcdm\ with Salpeter IMF.  The curves are those
corresponding to the optical depths $\tau(E_{\gamma})$
shown in Figure~\ref{fig:tau}.  The value of z$_s$
indicates the assumed source redshift.}
\label{fig:attenuation}
\end{figure}

Figure~\ref{fig:lcdmSalScalo} shows the EBL for the \lcdm\ model for the Scalo
and Salpeter IMFs.  As expected from our previous discussion, the predicted EBL
is much higher for the Salpeter IMF at both short and long wavelengths, but the
predictions are very similar in the 1-10 $\mu$m band that is most relevant for
$\sim$ TeV $\gamma$-ray attenuation from relatively nearby sources, in
agreement with MP96. Note that both EBL curves are consistent with the
lower limits from source counts at ultraviolet, optical, and near-infrared
wavelengths (filled symbols), but neither curve is high enough to agree with
the new DIRBE EBL detections at 140 and 240 $\mu$m \cite{dirbeEBL}.

The remaining results that we present are all for the Salpeter IMF.
Figure~\ref{fig:ebl_all} shows the predicted EBL for the four cosmological
models discussed above.  Note that the three models in which galaxy and star
formation are relatively early predict rather similar EBL, while CHDM predicts
generally lower EBL.  Figure~\ref{fig:evol} shows the reason for this in more
detail.  Only about 10\% of the EBL in the 1-10 $\mu$m band comes from $z\geq1$
for the CHDM model, compared to 20-40\% for the \lcdm\ model, in which galaxies
form considerably earlier.  (An alternative way of looking at the evolution of
the EBL is given in Figure 8 
of \cite{spblois}.) Keep in mind, however, that even in a model
with an ``early formation'' cosmology like \lcdm\ but
which assumes less efficient star formation at high redshift
(for example, the model with the star formation history indicated by
the light solid line in Figure~\ref{fig:sfrz}), the EBL will show a
steeper evolution than that shown in Figure~\ref{fig:evol} for LCDM.
So we expect some degeneracy between
cosmology and astrophysics.  However, to the extent
that the background cosmology may soon be determined by other methods,
measurements of the EBL will provide useful
constraints on the star formation history
of the Universe.

All of the models fall short of the DIRBE detection at 140 $\mu$m by at least
a factor of $\sim2.5$. This may be due to a number of effects that we have not
yet included in our modeling. Much of the far-infrared and sub-mm light is
probably produced by heavily extinguished ultra-luminous starburst
galaxies. The phenomenological work of Guiderdoni et al.~\cite{Guiderdoni98}
suggests that the contribution from this population may increase with
redshift. Our independent work suggests a physical reason for this: the galaxy
interactions that probably trigger these starburst events are more frequent at
higher redshift because of the higher density of the Universe, and galaxies are
more gas rich so the starburst events may be more dramatic
\cite{spf98}. Observations of nearby starburst galaxies (cf. \cite{calzetti})
suggest that the Galactic/SMC model that we have used here does not provide a
good description of dust extinction in actively star forming
galaxies. Additional contributions to the far-IR flux that we have not included
may come from AGN and from energy that is injected into the gas by supernovae
and later radiated at long wavelength. We will include the contribution of
starburst galaxies and address these other effects in improved models that we
are developing in collaboration with Bruno Guiderdoni and Julien
Devriendt. However, we do not expect that this improved treatment will have
much effect on the absorption of $\gamma$-rays with energies $\lsim 10$ TeV,
which we now discuss briefly.

\section{Attenuation of TeV Gamma Rays}

Here we have space to present only the results for one case, LCDM with
Salpeter IMF.  Figure~\ref{fig:tau} shows the optical depth of the universe
$\tau(E_{\gamma})$ as a function of $\gamma$-ray energy for varying
source redshifts, $z_s$.  The corresponding attenuation factors,
$\exp(-\tau)$, are shown in Figure~\ref{fig:attenuation}.
Note that for sources as near as Mrk 421 and 501 ($z=0.03$),
attenuation is predicted to be rather small between 1-10 TeV,
with little curvature in the spectrum.
Gamma-ray energies $E_{\gamma} \gsim 10$ TeV for local sources,
or sources at $z\gsim0.1$ for lower energies, will likely be
needed in order to see significant attenuation.

In \cite{bsmp} we will present results for several
cosmological models, and discuss dependence on IMF and star formation
prescriptions.  Bullock et al.~\cite{bullock} discusses the
difference in predicted $\gamma$-attenuation between Salpeter and
Scalo IMF for this same LCDM model.  The early universe is much more
transparent to 10-100 GeV $\gamma$-rays with the Scalo IMF, since the
fraction of high-mass stars is lower, and the ultraviolet flux density
is correspondingly reduced (cf. \cite{madauphinney,bsmp}).

\section{Conclusions}

\begin{itemize}
\item Semi-analytic models (SAMs) of galaxy formation provide a convenient and
powerful theoretical framework to determine how input assumptions --- e.g.,
cosmology, star formation history, and IMF --- affect the predicted
extragalactic background light (EBL) and the resulting $\sim$TeV $\gamma$-ray
attenuation.

\item The 1-10 $\mu$m EBL and the resulting attenuation of few-TeV
$\gamma$-rays reflect mainly the history of star formation in the
universe, with less attenuation for models such as CHDM in which
galaxies form relatively late.

\item The EBL at $\lsim 1$ $\mu$m and $\gsim 10$ $\mu$m is
significantly affected by the IMF and the modeling of the absorption
and reradiation by dust.
\item Gamma-ray energies 
$E_{\gamma} \gsim 10$ TeV and/or sources at
$z\gsim0.1$ will probably be needed to provide clear evidence of
attenuation due to $\gamma \gamma \rightarrow e^+e^-$.
\item Therefore, both space- and ground-based $\gamma$-ray telescopes will
be required to probe the spectra of AGNs at various redshifts, in
order to determine both
  \begin{itemize}
  \item the unabsorbed spectra, which will help determine how these
$\gamma$-rays are produced, and
  \item the intergalactic absorption, which as we have shown is
affected by cosmology, star formation history, IMF, and dust.
  \end{itemize}
\end{itemize}

\section*{Acknowledgments}

JRP and JSB were supported by NSF and NASA grants at UCSC, and RSS was
University Fellowship from The Hebrew University. JRP thanks Avishai
Dekel for hospitality at Hebrew University.  Donn MacMinn contributed
significantly to an early stage of this work, especially the dust
modeling.  His life was tragically cut short by a hit-and-run driver
as he was bicycling near Chicago on August 30, 1997.

\end{document}